\documentclass[aps,prl,twocolumn,showpacs,superscriptaddress,groupedaddress]{revtex4} 
\pdfoutput=1
\usepackage[version=3]{mhchem} 
\usepackage[dvips]{epsfig}
\usepackage{graphicx}  
\usepackage{dcolumn}   
\usepackage{bm}        
\usepackage{amssymb}   
\usepackage{color}
\usepackage{float}
\usepackage{hyperref}

\newcommand{\dif}{\text{d}}

\begin{document}

\title{Nonlinear optomechanical paddle nanocavities}

\author{Hamidreza Kaviani}
\affiliation{Institute for Quantum Science and Technology, University of Calgary, Calgary, AB, T2N 1N4, Canada}
\affiliation{National Institute for Nanotechnology, Edmonton, AB, T6G 2M9, Canada}

\author{Chris Healey}
\affiliation{Institute for Quantum Science and Technology, University of Calgary, Calgary, AB, T2N 1N4, Canada}
\affiliation{National Institute for Nanotechnology, Edmonton, AB, T6G 2M9, Canada}

\author{Marcelo Wu}
\affiliation{Institute for Quantum Science and Technology, University of Calgary, Calgary, AB, T2N 1N4, Canada}
\affiliation{National Institute for Nanotechnology, Edmonton, AB, T6G 2M9, Canada}

\author{Roohollah Ghobadi}
\affiliation{Institute of Atomic and Subatomic Physics, TU Wien, Stadionallee 2, 1020 Wien, Austria}

\author{Aaron Hryciw}
\affiliation{National Institute for Nanotechnology, Edmonton, AB, T6G 2M9, Canada}
\affiliation{nanoFAB Facility, University of Alberta, Edmonton, AB, T6G 2R3, Canada}

\author{Paul E. Barclay}
\email{pbarclay@ucalgary.ca}
\affiliation{Institute for Quantum Science and Technology, University of Calgary, Calgary, AB, T2N 1N4, Canada}
\affiliation{National Institute for Nanotechnology, Edmonton, AB, T6G 2M9, Canada}

\begin{abstract}

Nonlinear optomechanical coupling is the basis for many potential future experiments in quantum optomechanics (e.g., quantum non-demolition measurements, preparation of non-classical states), which to date have been difficult to  realize due to  small   non-linearity in typical optomechanical devices. Here we introduce an optomechanical system combining strong nonlinear optomechanical coupling, low mass and large optical mode spacing. This nanoscale ``paddle nanocavity''  supports mechanical resonances with hundreds of fg mass which couple nonlinearly to optical modes with a quadratic optomechanical coupling coefficient $g^{(2)} > 2\pi\times400$ MHz/nm$^2$, and a two phonon to single photon optomechanical coupling rate $\Delta \omega_0 > 2\pi\times 16$ Hz. This coupling relies on strong phonon-photon interactions in a structure whose optical mode spectrum is highly non--degenerate. Nonlinear optomechanical readout of thermally driven motion in these devices should be observable for T  $> 50 $ mK, and measurement of phonon shot noise is achievable. This shows that strong nonlinear effects can be realized without relying on coupling between nearly degenerate optical modes, thus avoiding parasitic linear coupling present in two mode systems.

\end{abstract}

\maketitle

The study of quantum properties of mesoscopic mechanical systems is a rapidly evolving field which has been propelled by recent advances in development of cavity optomechanical devices \cite{ref:aspelmeyer2013co}. Nanophotonic cavity optomechanical structures \cite{ref:eichenfield2009oc} allow co-localization of photons and femtogram to picogram mechanical excitations, and have enabled demonstrations of ultra-sensitive displacement and force detection \cite{ref:gavartin2012aho, ref:krause2012ahm, ref:liu2012wcs, ref:sun2012fdc, ref:wu2014ddo}, ground state cooling  \cite{ref:chan2011lcn} and optical squeezing \cite{ref:safavi2013sls}.  Development of cavity optomechanical systems with large nonlinear photon--phonon coupling has been motivated by quantum non-demolishing (QND) measurement of phonon number \cite{ref:thompson2008sdc} and  shot noise \cite{ref:clerk2010qmp}, as well as mechanical quantum state preparation \cite{ref:brawley2014nom}, study of photon--photon interactions \cite{ref:ludwig2012eqn}, mechanical squeezing and cooling \cite{ref:bhattacharya2008otc, ref:nunnenkamp2010csq, ref:biancofiore2011qdo}, and phonon-photon entanglement \cite{ref:liao2014spq}.

Recent progress in developing optomechanical systems with large nonlinear optomechanical coupling has been driven by studies of membrane-in-the-middle (MiM) \cite{ref:sankey2010stn, ref:flowers2012fcb, ref:karuza2013tlq, ref:lee2014mod} and whispering gallery mode \cite{ref:hill2013now, ref:doolin2014nos, ref:brawley2014nom} cavities. Demonstrations of massively enhanced quadratic coupling  \cite{ref:flowers2012fcb, ref:lee2014mod, ref:hill2013now} have exploited avoided crossings between nearly--degenerate optical modes, and have revealed rich multimode dynamics \cite{ref:lee2014mod}. To surpass bandwidth limits  \cite{ref:heinrich2010psz, ref:ludwig2012eqn} and {  parasitic linear coupling \cite{ref:miao2009sql} imposed by closely spaced optical modes, it is desirable to develop devices which combine strong nonlinear coupling and large optical mode spacing.  This can be achieved in short, low-mass, high-finesse optical cavities. In this work we present such a nanocavity optomechanical system, which couples modes possessing low optical loss and THz free spectral range, to mechanical resonances with femtogram mass, 300 kHz -- 220 MHz frequency, and large zero point fluctuation amplitude. This device has vanishing linear and large nonlinear optomechanical coupling, with quadratic optomechanical coupling coefficient $g^{(2)} \approx 2\pi\times 400$ MHz/nm$^2$ and single photon to two phonon coupling rate $\Delta \omega_0=2\pi\times 16$ Hz.}

The strength of photon--phonon interactions in nanocavity--optomechanical systems is determined by the modification of the optical mode dynamics via deformations to the nanocavity dielectric environment  from excitations of mechanical resonances.  In systems with dominantly dispersive optomechanical coupling, this dependence is expressed to second-order in mechanical resonance amplitude $x$ as  $\omega_o(x)=\omega_{0}+g^{(1)} x+\frac{1}{2} g^{(2)} x^{2}$, where $\omega_{o}$ is the cavity resonance frequency, and $g^{(1)}=\delta \omega_o/\delta x$, $g^{(2)}=\delta^2 \omega_o/\delta x^2$ are the first and second order optomechanical coupling coefficients. In nanophotonic devices, $x$ parameterizes a spatially varying modification to the local dielectric constant, $\Delta\epsilon(\mathbf{r};x)$, whose distribution depends on the mechanical resonance shape and is responsible for modifying the frequencies of the nanocavity optical resonances.

Insight into nonlinear optomechanical coupling in nanocavities is revealed by the dependence of $\delta\omega^{(2)}$ on the overlap between $\Delta\epsilon$ and the optical modes of the nanocavity \cite{ref:johnson2002ptm, ref:rodriguez2011bat}:
\begin{equation}
g^{(2)}=\frac{\omega}{2} \frac{|\langle E_{\omega}|\frac{\delta \epsilon}{\delta x}|E_{\omega}\rangle|^2}{|\langle E_{\omega}| \epsilon |E_{\omega}\rangle|^2} + \sum\limits_{\omega^{\prime}\neq\omega} g^{(2)}_{\omega^{\prime},\omega}.
\label{eq:second_order}
\end{equation}
where the first term is a ``self-term" and $g^{(2)}_{\omega^{\prime},\omega}$ represents cross--couplings between the fundamental mode of interest ($\omega$) and other modes supported by cavity ($\omega^{\prime}$):  
\begin{equation}
g^{(2)}_{\omega^{\prime},\omega}=-\left(\frac{\omega^{3}}{\omega^{\prime 2}-\omega^{2}}\right) \frac{|\langle E_{\omega^{\prime}}|\frac{\delta \epsilon}{\delta x} |E_{\omega}\rangle|^2}{\langle E_{\omega^{\prime}}| \epsilon |E_{\omega^{\prime}}\rangle \langle E_{\omega}| \epsilon |E_{\omega}\rangle}
\label{eq:overlap_terms}.
\end{equation}
Here $E_\omega$ denotes the electric field of a nanocavity mode at frequency $\omega$, and the inner product is an overlap surface integral defined in Ref.\ \cite{ref:johnson2002ptm} and developed in the context of optomechanics in Refs.\ \cite{ref:eichenfield2009mdc, ref:rodriguez2011bat} (see Supplementary information).  In cavity optomechanical systems with no linear coupling ($\delta\omega^{(1)} = 0$), the contribution in Eq.\ \eqref{eq:second_order} from the self--overlap of the dielectric perturbation vanishes, and the quadratic coupling is determined entirely by mechanically induced cross-coupling between the nanocavity's optical modes. Enhancing this coupling can be realized in two ways.  In the first approach, demonstrated in Refs.\ \cite{ref:sankey2010stn, ref:flowers2012fcb, ref:hill2013now, ref:karuza2013tlq, ref:lee2014mod}, the factor $\omega^{2}/(\omega^{\prime 2}-\omega^{2})$ can be enhanced in a cavity with nearly--degenerate modes ($\omega \sim \omega^{\prime}$) which are coupled by a mechanical perturbation. An alternative approach which is desirable to avoid multimode dynamics \cite{ref:lee2014mod} is to maximize the $g^{(2)}_{\omega^{\prime},\omega}$ overlap terms.  Here we investigate this route, and present a system with optical modes isolated by THz in frequency which possesses high quadratic optomechanical coupling owing to a strong overlap between optical and mechanical fields.

\begin{figure}
\centering
\includegraphics[width=1\columnwidth]{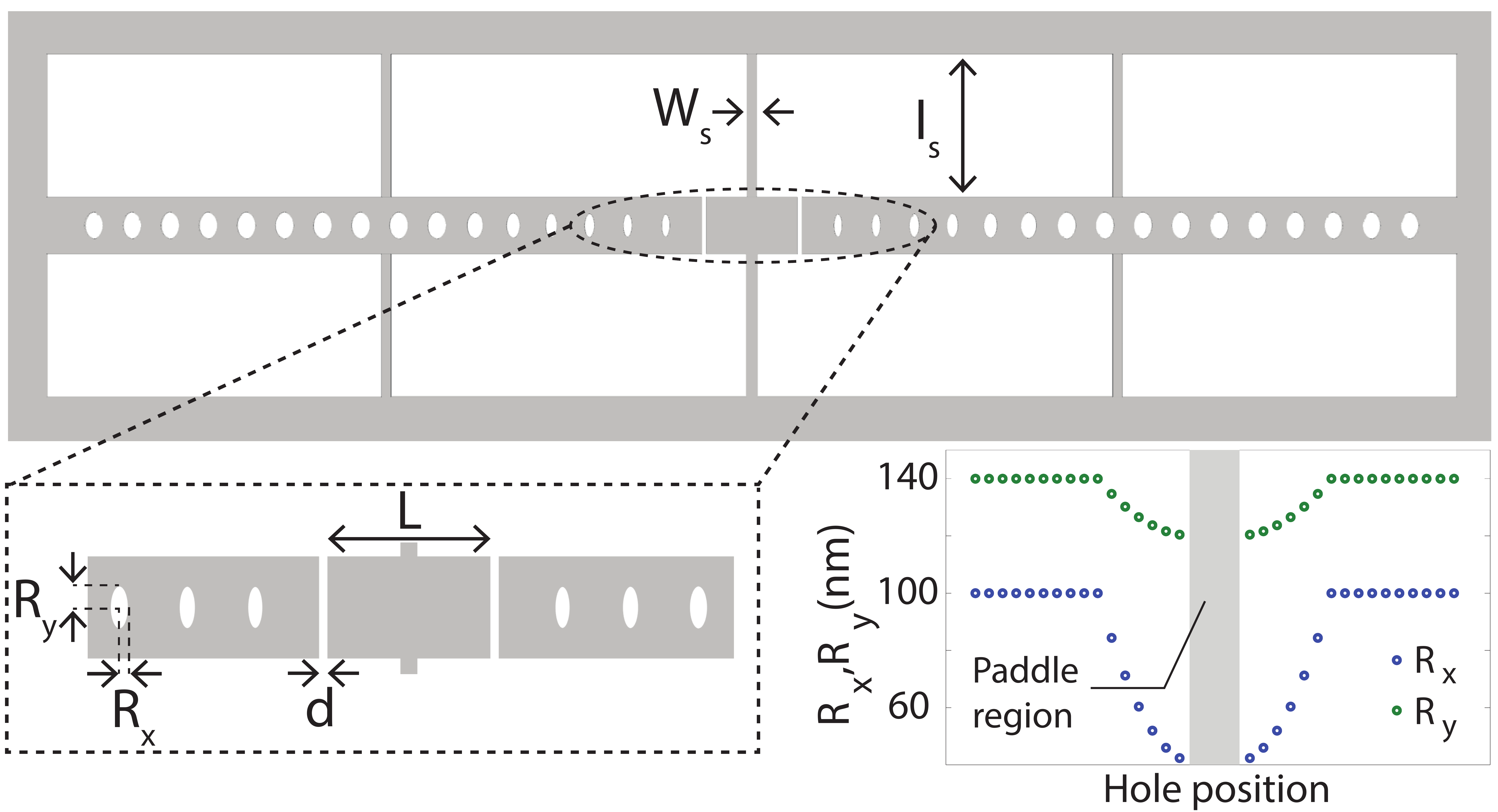}
\caption{Schematic of the photonic crystal paddle nanocavity (top-view).  The paddle is separated from photonic crystal nanobeam mirrors by gaps $d=50$ nm, and has length $L=958$ nm.  The elliptical hole horizontal and vertical semi-axes ($R_x, R_y$) are tapered as shown.}
\label{fig:schematic}
\end{figure}

\begin{figure}
\centering
\includegraphics[width=\columnwidth]{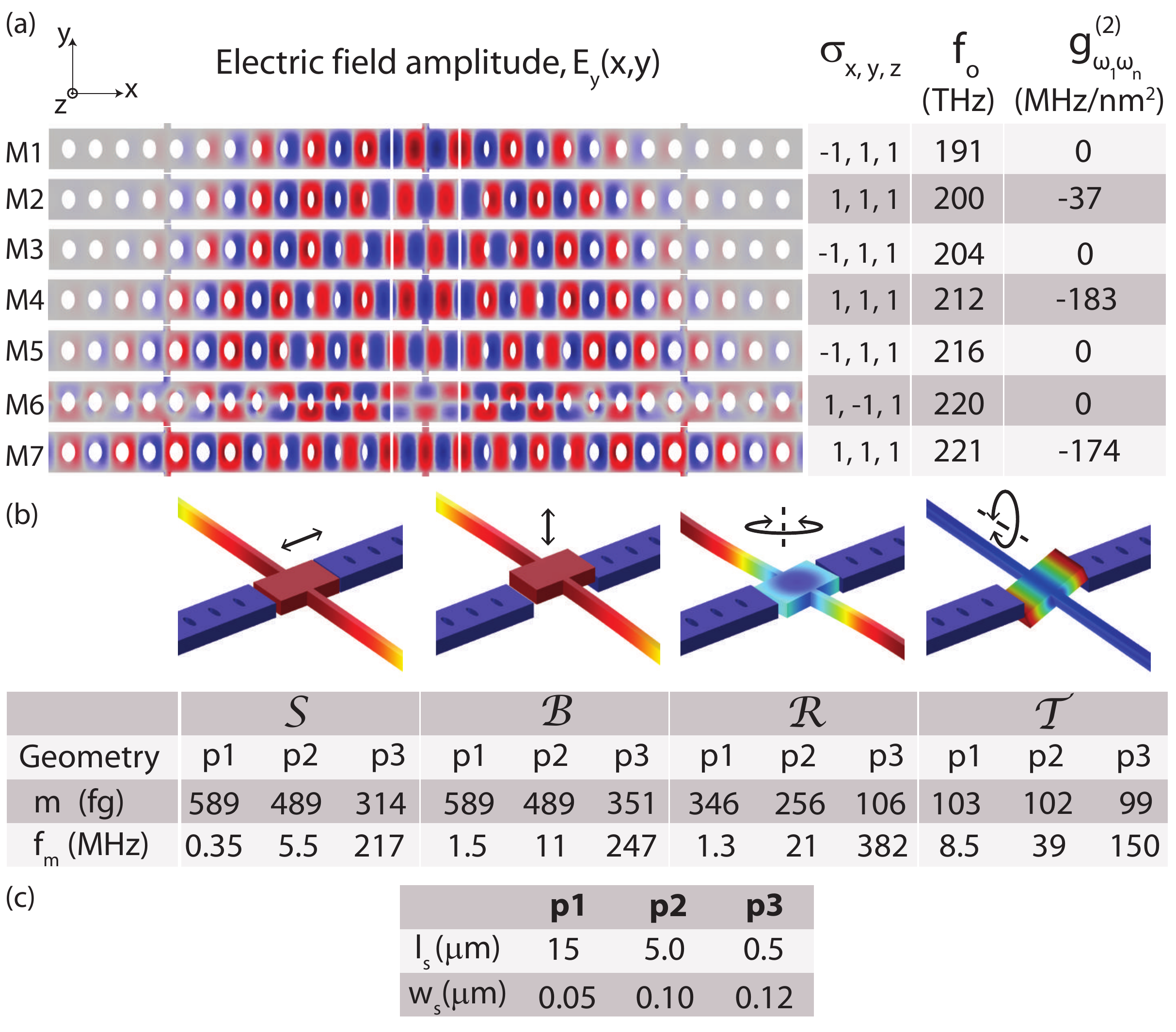}
\caption{  (a) Properties of the localized optical modes of the paddle nanocavity (labeled M1 -- M7): electric field distribution ($E_y$ component), spatial symmetry in $x-y$ plane, optical frequency, and contribution $g^{(2)}_{\omega_1 \omega_n}$ to the quadratic optomechanical coupling $g^{(2)}$ describing interaction between mode M1 and the $\cal{S}$ mechanical resonance shown in (b). (b) Displacement profiles and properties of the paddle nanocavity mechanical resonances. $m$ and $f_m$ are indicated for three support geometries, $p1-p3$, whose cross-sections are given in (c).}
\label{table}
\end{figure}

The optomechanical device studied here, illustrated in Fig.\ \ref{fig:schematic}, is a photonic crystal ``paddle nanocavity'' which combines operating principles of MiM cavities \cite{ref:thompson2008sdc, ref:jayich2008dom} and photonic crystal nanobeam optomechanical devices \cite{ref:eichenfield2009oc}. The device is designed to be fabricated from  silicon-on-insulator  (refractive index $n_\text{Si} = 3.48$, thickness $t = 220$ nm), and to support modes near $\lambda \sim 1550$ nm. A ``paddle'' element is suspended within the optical mode of the nanocavity defined by two photonic crystal nanobeam mirrors. The width of the gap ($d=50$ nm) separating the mirrors from the paddle is chosen for smooth variation in local effective-index of the structure \cite{ref:hryciw2013ods}, and the paddle length ($L=958$ nm) is set to $\approx 1.5 \lambda/n_\text{eff}$ \cite{ref:quan2010pcn}. This allows the nanocavity to support high optical quality factor ($Q_o$) modes. The length ($l_s$) and width ($w_s$) of the paddle supports can be adjusted to tailor its mechanical properties, although $l_s \ge 200$ nm and $w_s \le 200$ nm is required to not degrade $Q_o$.  We consider three support geometries, labeled $p1-p3$ (see \ref{table} for dimentions). All of these dimensions are realizable experimentally \cite{ref:wu2014ddo}. 

Figure \ref{table}(a) shows the { first seven} localized optical modes supported by the paddle nanocavity, calculated using finite element simulations (FEM) \cite{footnote}.  The lowest order mode ($M1$) has a resonance wavelength near 1550 nm ($\omega_o/2\pi = 191$ THz) and $Q_o > 1.3 \times 10^4$. The mechanical resonances of the paddle nanocavity were also calculated using FEM simulations, and the displacement profiles of the four lowest mechanical frequency resonances are shown in Fig.\ \ref{table}(b). They are referred to here as ``sliding'' ($\cal{S}$), ``bouncing'' ($\cal{B}$), ``rotational'' ($\cal{R}$) and ``torsional'' ($\cal{T}$) resonances. As discussed below, we are particularly interested in the $\cal{S}$ resonance, whose frequency and effective mass \cite{ref:eichenfield2009mdc} varies between $f_m = 0.35 - 217\,$MHz and $m = 314-589\,$fg for the support geometries $p1-p3$, as described in Fig.\ \ref{table}(b). Appropriate selection of geometry $p1-p3$ depends on the application, with $p1$ suited for sensitive actuation, $p2$ a compromise between ease of fabrication and sensitivity, and $p3$ for high frequency operation and low thermal phonon occupation.

The spatial symmetry of the nanocavity results in vanishing $g^{(1)}$ for the mechanical resonances considered here. The intensity $E^2(x,y,z)$ of each nanocavity optical mode has even symmetry, denoted $\sigma_{x,y,z} =1$, while the mechanical resonances induce perturbation $\Delta\epsilon$ which is odd in at least one direction, characterized by $\sigma_{x} = -1$ ($\cal{S},\cal{R}$),  $\sigma_{y} = -1$ ($\cal{R}$) or $\sigma_{z} = -1$ ($\cal{B},\cal{T}$).  As a result, $g^{(1)} \propto \langle E_{\omega}|\delta \epsilon|E_{\omega}\rangle = 0$. Similarly, the second order self--overlap term in Eq.\ \eqref{eq:second_order} is also zero. However, the electric field amplitude $E(x,y,z)$ may be even or odd, resulting in non--zero cross-coupling $g^{(2)}_{\omega^{\prime},\omega}$ between optical modes with opposite $\sigma_{x,y,z}$.  For example, displacement in the $\hat{x}$ direction of $\cal{S}$ couples optical modes with opposite $\sigma_x$. In contrast,  displacement in the $\hat{z}$ direction by the $\cal{B}$ and $\cal{T}$ resonances does not induce cross--coupling, as the localized optical modes all have even vertical symmetry ($\sigma_z = 1$). Here we focus on the nonlinear coupling between the $\cal{S}$ resonance and the $M1$ mode of the nanocavity.

To evaluate $g^{(2)}_{\omega,\omega'}$, the mechanical and optical field profiles were input into Eq.\ \eqref{eq:overlap_terms}.  The resulting contributions $g_{\omega_1 \omega_n}^{(2)}$ of each localized mode to $g^{(2)}$ for optomechanical coupling between the $\cal{S}$ resonance and the $M1$ mode are summarized in Fig.\ \ref{table}(a). Contributions from delocalized modes are neglected due to their large mode volume and low overlap. The imaginary part of $\omega_o$, which is small for the localized modes whose $Q_o > 10^2$, is also ignored. A total $g^{(2)}/2\pi \approx -400$ MHz/nm$^{2}$ is predicted, which matches with our direct FEM calculations (see Supplementary information). The corresponding single photon to two phonon coupling rate, $\Delta\omega_o$ depends on the support geometry.  For the most flexible $p1$ geometry, $\Delta\omega_{o}\equiv|g^{(2)}x_{zpf}^{2}|=2\pi\times 16$ Hz, where $x_{zpf}=\sqrt{{\hbar}/{2 m \omega_m}}$. { This $\Delta\omega_{o}$ is about four orders of magnitude higher than typical MiM systems} \cite{ref:sankey2010stn, ref:lee2014mod}, while the mode spacing is five orders of magnitude higher than other nonlinear optomechanical systems \cite{ref:sankey2010stn, ref:ludwig2012eqn, ref:hill2013now}. The dominant contributions to $g^{(2)}$ arise from cross--coupling between modes $M1\leftrightarrow M4$ and $M1\leftrightarrow M7$ due to strong spatial overlap between their fields and the paddle--nanobeam gaps.  Increasing $g^{(2)}$ through additional optimization, for example by concentrating the optical field more strongly in the gap, should be possible.  

Given $g^{(2)}$ of the paddle nanocavity, the optical response of the device can be predicted.  In experimental applications of optomechanical nanocavities, photons are coupled into and out of the nanocavity using an external waveguide.  Mechanical fluctuations, $x(t)$ are monitored via variations, $dT(t)$, of the waveguide transmission, $T$. In the sideband unresolved regime ($\omega_m \ll \omega_o/Q_o$), optomechanical coupling  results in a fluctuating waveguide output
$dT=G_1 x(t)+\frac{1}{2} G_2 x(t)^2$, where 
\begin{align}
G_1 &= \frac{dT}{dx}= g^{(1)} \frac{dT}{d\Delta},\\
G_2 &= \frac{d^2 T}{dx^2}= g^{(2)} \frac{dT}{d\Delta} + \left(g^{(1)}\right)^{2} \frac{d^{2}T}{d\Delta^{2}}. \label{eq:G2}
\end{align} 
Here $\Delta = \omega - \omega_o$ is the detuning between input photons and the nanocavity mode, and $dT/d\Delta$, $d^2T/d\Delta^2$ are the slope and curvature of the Lorentzian cavity resonance in $T(\Delta)$. { Eq.\ \eqref{eq:G2} shows that in general, both nonlinear transduction of linear optomechanics and linear transduction of nonlinear optomechanics contribute to the second order signal}. The nonlinear mechanical displacement can be measured through photodetection of the waveguide optical output. For input power $P_i$, the waveguide output optical power spectral density (PSD) due to transduction of $x^2(t)$ is $S_{P}^{(2)}(\omega) = \frac{1}{4}P_i^2 G_2^{2} S_{x^{2}}(\omega)$,
where  $S_{x^{2}}(\omega)$ is the PSD of the $x^2$ mechanical motion of the mechanical resonance.  To analyze the possibility of observing this signal, it is instructive to consider the scenario of a thermally--driven mechanical resonance.  As shown in the Supplementary information and Refs. \cite{ref:hauer2015nps, ref:nunnenkamp2010csq}, $S_{x^{2}}(\omega)$ of a resonator in a $\bar{n}$ phonon number thermal state is
\begin{align}
\nonumber S_{x^2}(\omega)=2 x_{zpf}^4&\left(\frac{2 \Gamma (\bar{n}+1)^2}{\Gamma^2+(\omega-2\omega_m)^2}+\frac{2 \Gamma \bar{n}^2}{\Gamma^2+(\omega+2\omega_m)^2}\right.\\&+\left.\frac{8\Gamma \bar{n}(\bar{n}+1)+1}{\Gamma^2+\omega^2}\right),
\label{S_x2x2}
\end{align}
where $\Gamma = \omega_m/Q_m$ and $Q_m$ is the mechanical quality factor. 

\begin{figure}
\includegraphics[width=0.91\columnwidth]{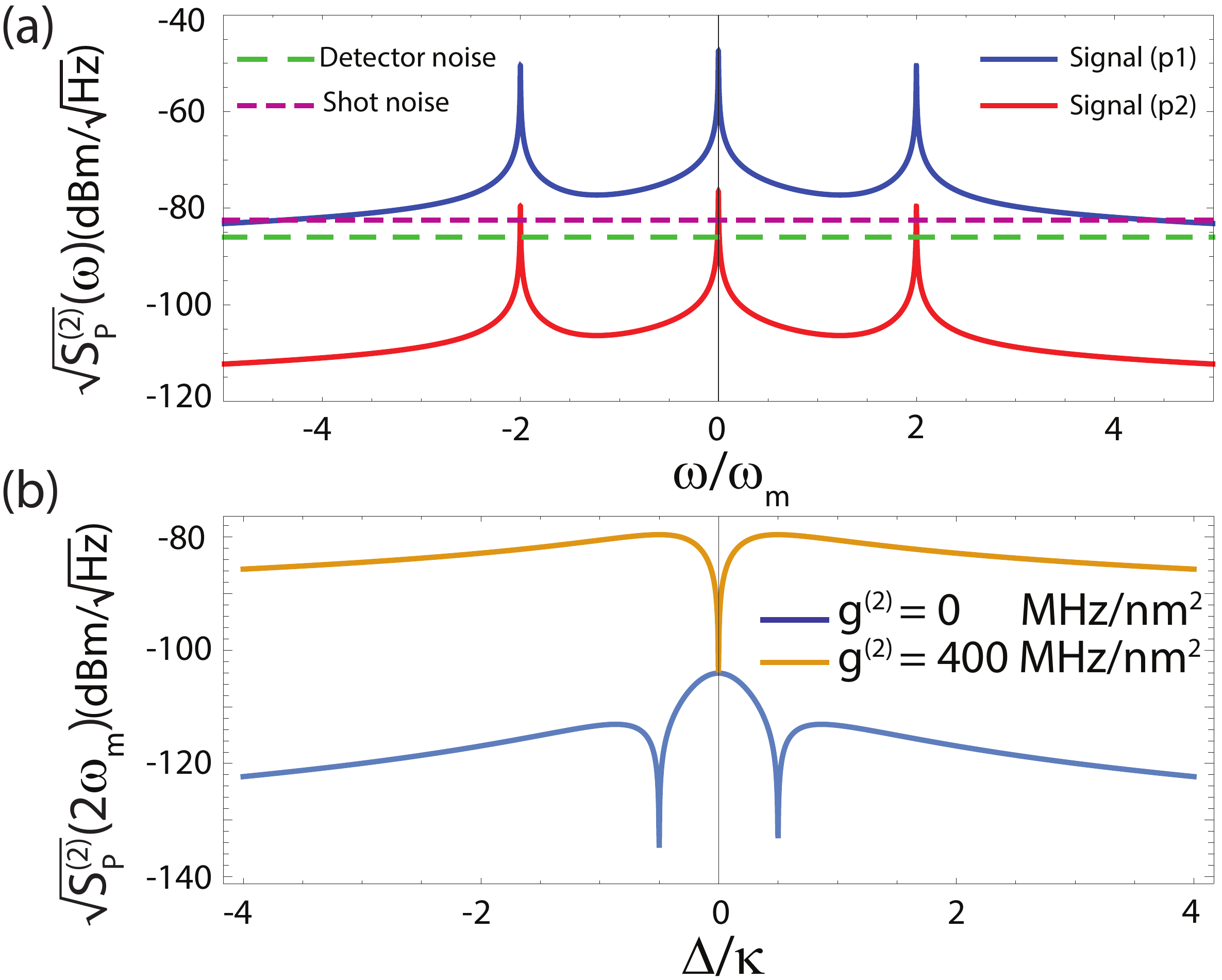}
\caption{(a) $S_{P}^{(2)}(\omega)$ generated by thermal motion of the $\cal{S}$ mode of a paddle nanocavity for $p1$ and $p2$ support geometry, assuming room temperature operation, $\Delta=\kappa/2$,  $P_i = 100\,\mu$W and $Q_m = 10^3$. (b) $S_{P}^{(2)}(2 \omega_{m})$ as a function of detuning $\Delta$, for varying quadratic coupling strengths $g^{(2)}$.}
\label{Mechanical_Signal}
\end{figure}

Figure \ref{Mechanical_Signal} shows $S_P^{(2)}(\omega)$ predicted from Eq.\ \eqref{S_x2x2}, for the $\cal{S}$ mode of a paddle nanocavity at room temperature ($T_b = 300$K).  The input optical field is set to $P_{i}=100$ $\mu$W, with detuning $\Delta=\kappa/2$ to maximize the nonlinear optomechanical coupling contribution. The predicted $S_P^{(2)}(\omega)$ is shown for $p1$ and $p2$ support geometries, assuming $Q_{m}=10^3$, $Q_{o}=1.4 \times 10^4$,  and relatively weak fiber coupling $T_o=0.90$. Note that $Q_m$ is specified assuming the device is operating in moderate vacuum conditions \cite{ref:wu2014ddo}, and can increase to $10^{5}$ in cryogenic vacuum conditions \cite{ref:safavi2013sls}. Also shown are estimated noise levels, assuming direct photodetection using a Newport 1811 photoreceiver (NEP=$2.5$ pW/$\sqrt{Hz}$). Resonances in $S_P^{(2)}$ are evident at $\omega = 2\omega_m$ and $\omega=0$, corresponding to energies of the two phonon processes characteristic of $x^2$ optomechanical coupling. Figure \ref{Mechanical_Signal} suggests that even for these relatively modest device parameters, the nonlinear signal at $2\omega_m$ is observable. This signal can be further enhanced with improved device performance. For example, if $Q_m = 10^4$, the nonlinear signal is visible for temperatures as low as 50 mK for the $p1$ geometry. Note that additional technical noise will increase as $f_m$ is further decreased into the kHz range.

Nonlinear optomechanical coupling can be differentiated from nonlinear transduction by the $\Delta$ dependence of the nonlinear signal. This is demonstrated in Fig.\ \ref{Mechanical_Signal}, which shows $S_P^{(2)}(2\omega_m, \Delta)$ with and without quadratic coupling, assuming that fabrication imperfections introduce nominal $g^{(1)}/2\pi=50$ MHz/nm.  This demonstrates that at $\Delta = \kappa/2$, the nonlinear signal is dominantly from nonlinear optomechanical coupling.

Next we study the feasibility of QND phonon measurement using a paddle nanocavity. High $\omega_m$ is advantageous for ground state cooling which is required for QND measurements. The large optical mode spacing of the paddle nanocavity allows this without introducing Zener tunneling effects \cite{ref:ludwig2012eqn} { or parasitic linear coupling and resulting backaction \cite{ref:miao2009sql}}. Cryogenic temperature of $10$ mK could directly cool the $\cal{S}$ resonance of the $p3$ structure to its quantum ground state.  For feasible optical and mechanical quality factors $Q_{o}=10^{6}$ \cite{ref:hryciw2013ods, ref:chan2011lcn} and $Q_{m}=10^{5}$ \cite{ref:safavi2013sls}, the signal to noise ratio (SNR) introduced in Ref.\ \cite{ref:thompson2008sdc} of a quantum jump measurement in such a device is $\Sigma^{(0)} = \tau^{(0)}_{tot} \Delta \omega_{0}^{2}/S_{\omega_{o}}=6.4\times10^{-8}$.  Here $\tau^{(0)}_{tot}$ is the thermal lifetime quantifying the rate of decoherence due to bath phonons of the ground state cooled nanomechanical resonator, and $S_{\omega_{o}}$ is the shot noise limited sensitivity of an ideal Pound-Drever-Hall detector.  Introducing laser cooling would potentially allow preparation of the $p1$ device to its quantum ground state, where the larger $x_{zpf}$ and $\Delta\omega_o$ { increases} $\Sigma^{(0)}$ { to} ${2.1\times 10^{-5}}$.  However, this would require development of sideband unresolved nonlinear optomechanical cooling \cite{ref:nunnenkamp2010csq}.

A more feasible approach for observing discreteness of the paddle nanocavity mechanical energy is a QND measurement of phonon shot noise \cite{ref:clerk2010qmp}. The SNR of such a measurement scales with the magnitude of an applied drive, which  enhances the signal by $S=8n_d \bar{n}\Sigma^{(0)}$, where $n_d$ is the drive amplitude in units of phonon number, and $\bar{n} < 1$ for a resonator in the quantum ground state. Using a $p3$ structure, SNR of above one is achievable assuming a drive amplitude of $62$ pm ($n_d\approx 7.8\times10^{6}$) and thermal bath phonon number $\bar{n}=1/4$. 

In conclusion, we have designed a single--mode nonlinear optomechanical nanocavity with THz mode spacing. The quadratic optomechanical coupling coefficient $g^{(2)}/2\pi=400$ MHz/nm$^{2}$ and single photon to two phonon coupling rate $\Delta\omega_0 /2\pi= 16$ Hz of this system are among the largest single--mode quadratic optomechanical couplings predicted to-date. Observing a thermal nonlinear signal from this structure is possible in realistic conditions, and a continuous QND measurements of phonon shot noise may be achievable for optimized device parameters.

\section*{Funding Information}
Natural Science and Engineering Research Council of Canada, Canada Foundation for Innovation, Alberta Innovates Technology Futures, WWTF, Austrian Science Fund (FWF) through SFB FOQUS and the START grant Y 591-N16. 
\\
\section*{Acknowledgments}
We would like to thank David Lake for helpful discussions.

\pagebreak
\widetext
\begin{center}
\textbf{\large Supplementary Information for ``Nonlinear optomechanical paddle nanocavities''}
\end{center}
\setcounter{equation}{0}
\setcounter{figure}{0}
\setcounter{table}{0}
\setcounter{page}{1}
\makeatletter
\renewcommand{\theequation}{S\arabic{equation}}
\renewcommand{\thefigure}{S\arabic{figure}}
\renewcommand{\bibnumfmt}[1]{[S#1]}
\renewcommand{\citenumfont}[1]{S#1}
\vspace{40 pt}

\section{Evaluation of the nonlinear optomechanical coupling coefficient}

The matrix element used in the perturbation theory calculation of $g^{(2)}$ is a measure of the overlap of the nanocavity optical fields and the shifting dielectric boundaries of the mechanical resonance.  It is discussed in detail in Refs.\ \cite{sref:johnson2002ptm, sref:eichenfield2009mdc, sref:rodriguez2011bat}, and is given by 
\begin{equation}
\langle E_{\omega^{\prime}}|\frac{\delta \epsilon}{\delta x}|E_{\omega}\rangle = \int \dif A ({\bf q\cdot\hat{n}}) \left[ \Delta \epsilon \, \textbf{e}^{\parallel}_{\omega^{\prime}} \cdot\textbf{e}^{*\parallel}_\omega - \Delta (\epsilon^{-1})\, \textbf{d}^{\perp}_{\omega^{\prime}} \cdot  \textbf{d}^{*\perp}_\omega \right]
\end{equation}
where the integral is evaluated over the surface of the nanocavity, and $\textbf{e}^{\parallel}_\omega$ and $\textbf{d}^{\perp}_\omega$ are the components of the the optical mode electric and displacement fields parallel and perpendicular to the surface, respectively. The perturbation introduced by the mechanical resonance is described by the normalized displacement of the dielectric boundaries, $\textbf{q}=\textbf{Q(r)}/|\textbf{Q(r)}|_\text{max}$ where $\textbf{Q}(r)$ is the vectorial displacement field. For the device studied here, the dielectric contrast is constant, and is described by $\Delta \epsilon=\epsilon_1-\epsilon_2$ and $\Delta (\epsilon^{-1})=1/\epsilon_2-1/\epsilon_1$ , where $\epsilon_1$ is the dielectric constant of the nanocavity, and $\epsilon_2 = 1$ is the dielectric constant of the surrounding medium.

\section{Nonlinear optomechanical signal}

Here we analyze the optical power spectrum generated by a thermally driven mechanical oscillator quadratically coupled to an optical nanocavity. As described by Eq.\ (5) in the main text, the optical energy spectrum of a quadratically coupled mechanical resonator in a cavity optomechanical system can be written in terms of the autocorrelation of displacement squared,
\begin{equation}
S_{x^2}(\omega)= \int_{-\infty}^{+\infty} \langle x^2(t) x^2(0) \rangle e^{-i \omega t} dt.
\label{eq:S_xx}
\end{equation}   
Expressing the displacement in terms of annihilation and creation operators $b$ and $b^{\dagger}$ as $x=x_{zpf}(b e^{i \omega_{m} t}+b^{\dagger} e^{-i \omega_{m} t})$ and substituting the displacement operator into Eq.\ \eqref{eq:S_xx} yields
\begin{align}
\nonumber S_{x^2}(&\omega)=2 \pi x_{zpf}^{4} \left[( 2(\bar{n}+1)^2 \delta (\omega-2 \omega_m)\right.\\ &+\left.2\bar{n}^2 \delta (\omega+2 \omega_m) + (8 \bar{n}(\bar{n}+1)+1) \delta (\omega)\right].
\end{align}
where $\bar{n}$ is the mean thermal phonon number and $T_b$ is the bath temperature. { For large phonon numbers $\bar{n} \gg 1$, it is approximated by $\bar{n}=k_b T_b/\hbar \omega_{m}$ and the area under the nonlinear spectrum is given by}  
\begin{equation}
 \int^{+\infty}_{-\infty} S_{x^2} \frac{d\omega}{2 \pi}= 12 \bar{n} x_{zpf}^{4}= 3 \left(\frac{k_b T_b}{m \omega_m^2}\right)^2 
 \label{eq:Normalization}
 \end{equation}
which is in agreement with the moment relation for a thermal distribution $\langle x^4 \rangle=3 \langle x^2 \rangle^2$ \cite{sref:kardar2007spp}.
For low loss mechanical resonators ($\Gamma \ll \omega_m$) we can { replace the delta functions with a Lorentzian} $\delta(\omega-\omega_{m})=\frac{1}{\pi} \frac{\Gamma} {\Gamma^2+(\omega-\omega_m)^2}$, resulting in the following formula for power spectral density,
\begin{align}
\nonumber S_{x^2}(\omega)=2 x_{zpf}^4 &\left(\frac{2\Gamma (\bar{n}+1)^2}{\Gamma^2+(\omega-2\omega_m)^2}+\frac{2\Gamma \bar{n}^2}{\Gamma^2+(\omega+2\omega_m)^2}\right. \\
&\left.+\frac{8\Gamma \bar{n}(\bar{n}+1)+1}{\Gamma^2+\omega^2}\right).
\label{eq:quantum S_xx}
\end{align}
 Assuming a large thermal phonon occupancy ($\bar{n} \gg 1$), for frequencies near the double mechanical frequency ($\omega \approx 2 \omega_m$) we obtain following normalized form (using Eq.\ \eqref{eq:Normalization}) for nonlinear power spectral density
\begin{align}
\nonumber S_{x^2}(\omega) &= 96\left(\frac{k_B T_b}{m}\right)^2 \frac{\omega_m}{Q_m} \times \\ &{\frac{1}{[(\omega^2-4\omega_m^2)^2+(\frac{2 \omega_m \omega}{Q_m})^2] [\omega^2+(\frac{\omega_m}{2 Q_m})^2]}}
\label{eq:quantum_signal_limit}
\end{align}

One obtains a similar result from a classical analysis which assumes that during the mechanical decay time, $\Delta t \approx 1/\Gamma$, the thermal force acts as a delta function ``kick''. In this approximation, the spectral density of the thermal force is given by \cite{sref:saulson1990tnm} 
\begin{equation}
S_{FF}(\omega)=\frac{|F(\omega)|^2}{\Delta t}= \frac{2 k_B T_b m \omega_m}{Q_m}
\end{equation}
For a measurement time on the order of $\Delta t$, 
\begin{equation}
F(\omega)=\sqrt{\frac{2 k_B T_b m\omega_m \Delta t}{Q_m}} 
\end{equation}
and 
\begin{equation}
x(\omega)=\sqrt{\frac{2 k_B T_b  \omega_m \Delta t}{m Q_m}} \frac{1}{\omega_m^2-\omega^2+i\Gamma \omega}
\label{eq:x}
\end{equation}

\begin{figure}[H]
\centering
\includegraphics[width=0.7\columnwidth]{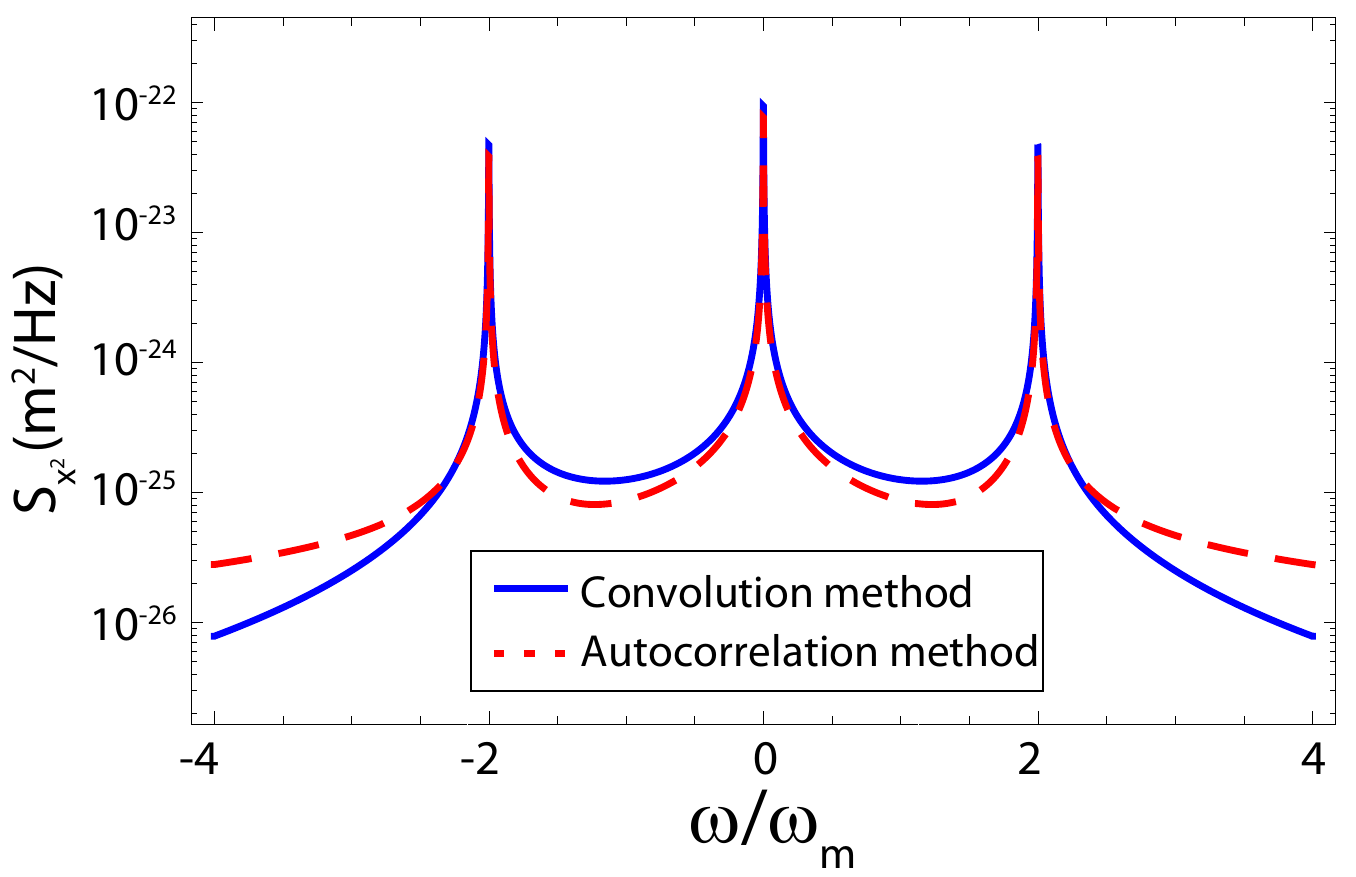}
\caption{Comparison of the power spectral density of the nonlinear signal obtained from exact autocorrelation analysis resulting in expression Eq.\ \eqref{eq:quantum S_xx} (red) and the approximate thermal analysis resulting in Eq.\ \eqref{eq:classical_signal} (blue), assuming $\omega_m=2\pi\times5.5$ MHz, $Q_m=1000$, and $T_b = 300$K.}
\label{fig:1}
\end{figure}

From  Eq.\ \eqref{eq:x} and Eq.\ \eqref{eq:S_xx}, and using the convolution properties of Fourier transforms, we find 
\begin{align}
\nonumber S_{x^2}&= 96\left(\frac{k_B T_b}{m}\right)^2 \frac{\omega_m}{Q_m} \times \\ &{\frac{1}{[(\omega^2-4\omega_m^2)^2+(\frac{2 \omega_m \omega}{Q_m})^2] [\omega^2+(\frac{\omega_m}{2 Q_m})^2]}},
\label{eq:classical_signal}
\end{align}
 after imposing the normalization given by Eq.\ \eqref{eq:Normalization}. As illustrated in Fig.\ \ref{fig:1},  the classical nonlinear signal described by Eq.\ \eqref{eq:classical_signal}  matches the quantum result of Eq.\ \eqref{eq:quantum S_xx} when $\bar{n} \gg 1$, in the neighbourhood of $\omega \sim 2 \omega_m$. { This analysis is in agreement with results in Ref.\ \cite{sref:hauer2015nps}}. 
 
\section{Validating the second order perturbation theory}

The accuracy of the second order perturbation theory, whose use has not been previously reported for nanophotonic cavity--optomechanical devices to the best of our knowledge, was tested by comparing its results with FEM calculations of $\omega_o(x)$, Here $x$ parameterizes the paddle displacement from the center position between the two mirrors of the simulated structure. This displacement closely approximates the motion of the $\cal{S}$ resonance which we are primarily interested in here.

\begin{figure}
\centering
\includegraphics[width=0.7\columnwidth]{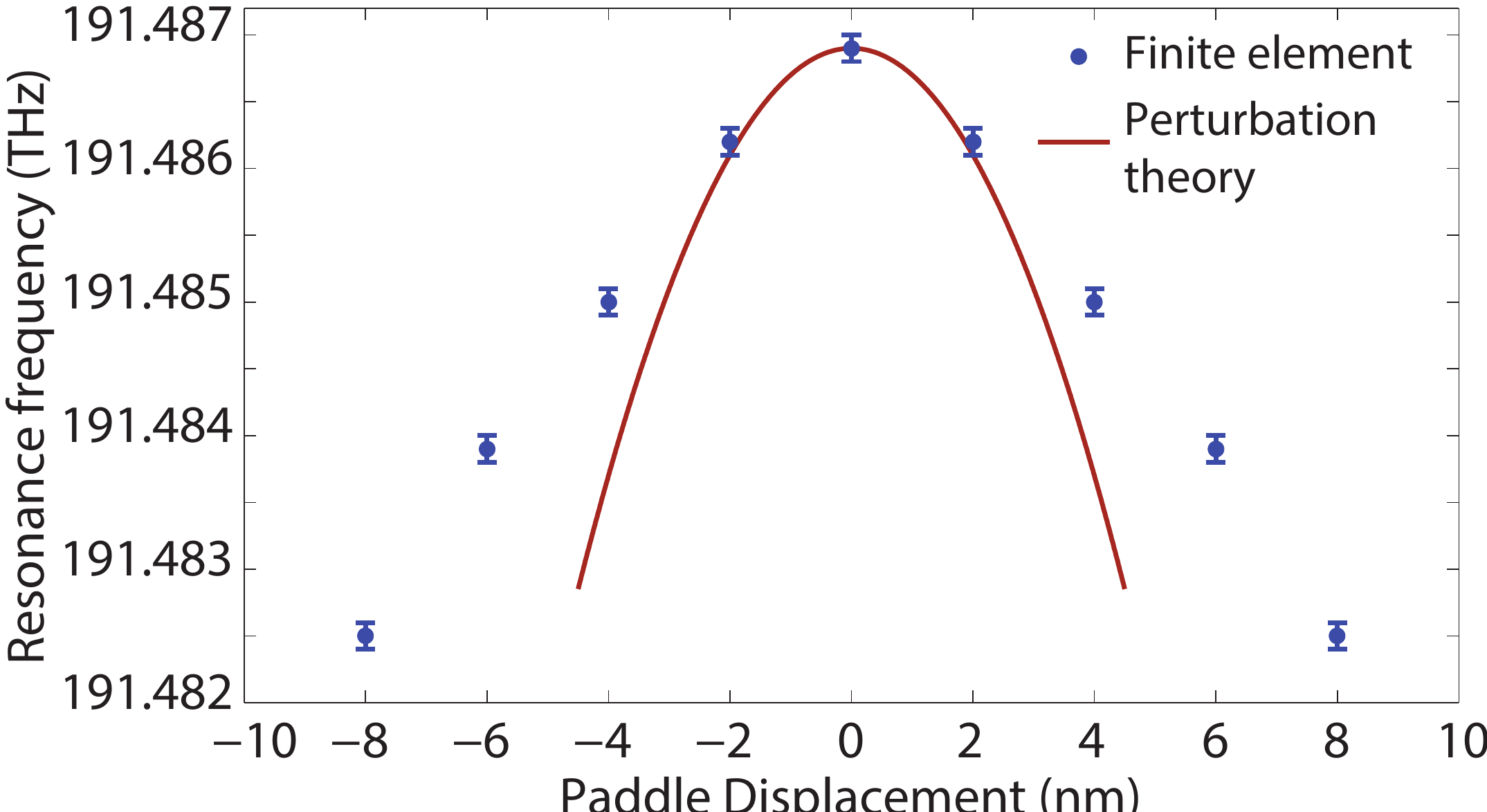}
\caption{Frequency of the $M1$ nanocavity mode  as a function of paddle displacement along the $x$-axis, calculated using FEM simulations of $\omega_o(x)$ (data points) and perturbation theory prediction $\omega_o(x) = \omega_o(0) + (1/2)g^{(2)}x^2$ (solid line) { with $g^{(2)}/2\pi=400\,\text{MHz/nm}^2$}. Error bars determined by the lowest significant digit of $\omega_o$.}
\label{f_x}
\end{figure}

This comparison is shown in Fig.\ \ref{f_x}, where we find good agreement for displacements $|x| \le 2$ nm, and deviation for larger displacements as the perturbation condition breaks down. This agreement confirms the validity of the assumptions underlying the second order perturbation theory. It also highlights the suitability of this method, as extracting  $g^{(2)}$ from parameterized FEM simulations has considerable uncertainty due a $2$ nm minimum mesh available with our computation tool.

\begin{thebibliography}{10}
\bibitem{footnote}
COMSOL software was used for all FEM simulations.
\end{thebibliography}
\end{document}